\documentstyle[12pt]{article}

\expandafter\ifx\csname mathrm\endcsname\relax\def\mathrm#1{{\rm #1}}\fi

\makeatletter
\if@twoside \oddsidemargin -17pt \evensidemargin 00pt
\else \oddsidemargin 00pt \evensidemargin 00pt
\fi
\expandafter\ifx\csname mathrm\endcsname\relax\def\mathrm#1{{\rm #1}}\fi

\makeatother

\def\beq{\begin{equation}}
\def\eeq{\end{equation}}
\def\beqar{\begin{eqnarray}}
\def\eeqar{\end{eqnarray}}
\def\barr#1{\begin{array}{#1}}
\def\earr{\end{array}}
\def\bfi{\begin{figure}}
\def\efi{\end{figure}}
\def\btab{\begin{table}}
\def\etab{\end{table}}
\def\bce{\begin{center}}
\def\ece{\end{center}}
\def\nn{\nonumber}

\def\text{\textstyle}

\def\eps{\varepsilon}

\def\si{\sigma}

\def\Ga{\Gamma}
\def\De{\Delta}

\def\refeq#1{\mbox{(\ref{#1})}}
\def\reffi#1{\mbox{Fig.~\ref{#1}}}

\def\refta#1{\mbox{Tab.~\ref{#1}}}
\def\citere#1{\mbox{Ref.~\cite{#1}}}
\def\citeres#1{\mbox{Refs.~\cite{#1}}}

\def\solid{\raise.9mm\hbox{\protect\rule{1.1cm}{.2mm}}}
\def\dash{\raise.9mm\hbox{\protect\rule{2mm}{.2mm}}\hspace*{1mm}}

\newcommand{\GeV}{\unskip\,\mathrm{GeV}}
\newcommand{\MeV}{\unskip\,\mathrm{MeV}}
\newcommand{\TeV}{\unskip\,\mathrm{TeV}}

\def\mathswitchr#1{\relax\ifmmode{\mathrm{#1}}\else$\mathrm{#1}$\fi}

\newcommand{\PW}{\mathswitchr W}
\newcommand{\PZ}{\mathswitchr Z}

\newcommand{\PH}{\mathswitchr H}

\newcommand{\Pd}{\mathswitchr d}

\newcommand{\Pu}{\mathswitchr u}

\newcommand{\Pb}{\mathswitchr b}
\newcommand{\Pc}{\mathswitchr c}
\newcommand{\Pt}{\mathswitchr t}

\newcommand{\Zbb}{$\PZ\to\Pb\bar\Pb$}
\newcommand{\Zcc}{$\PZ\to\Pc\bar\Pc$}

\newcommand{\Rb}{R_\Pb}
\newcommand{\Rc}{R_\Pc}
\newcommand{\Gb}{\Ga_\Pb}
\newcommand{\Gc}{\Ga_\Pc}

\newcommand{\Gh}{\Ga_{\mathrm h}}
\newcommand{\GT}{\Ga_{\mathrm T}}
\newcommand{\Gl}{\Ga_{\mathrm l}}

\def\mathswitch#1{\relax\ifmmode#1\else$#1$\fi}

\newcommand{\MW}{\mathswitch {M_\PW}}

\newcommand{\MZ}{\mathswitch {M_\PZ}}
\newcommand{\MH}{\mathswitch {M_\PH}}

\newcommand{\Mb}{\mathswitch {m_\Pb}}
\newcommand{\Mt}{\mathswitch {m_\Pt}}

\newcommand{\scrs}{\scriptscriptstyle}

\newcommand{\swbar}{\mathswitch {\bar s_{\scrs\PW}}}

\newcommand{\GF}{\mathswitch {G_\mu}}

\newcommand{\chidof}{\chi^2_{\mathrm{min}}/_{\mathrm{d.o.f.}}}
\newcommand{\nolimit}{{\tiny\gsim}1000}
\newcommand{\yb}{y_\Pb}

\newcommand{\yu}{y_\Pu}
\newcommand{\yd}{y_\Pd}
\newcommand{\yc}{y_\Pc}

\newcommand{\alpz}{\alpha(\MZ^2)}
\newcommand{\alps}{\alpha_{\mathrm s}}
\newcommand{\alpsz}{\alpha_{\mathrm s}(\MZ^2)}

\newcommand{\yh}{y_{\mathrm h}}

\newcommand{\QCD}{{\mathrm{QCD}}}
\newcommand{\QED}{{\mathrm{QED}}}
\newcommand{\LEP}{{\mathrm{LEP}}}
\newcommand{\SLD}{{\mathrm{SLD}}}
\newcommand{\SM}{{\mathrm{SM}}}

\def\draftdate{\relax}
\def\mda{\relax}
\def\mua{\relax}
\def\mla{\relax}
\def\draft{
\def\thtystars{******************************}
\def\sixtystars{\thtystars\thtystars}
\typeout{}
\typeout{\sixtystars**}
\typeout{* Draft mode!
         For final version remove \protect\draft\space in source file *}
\typeout{\sixtystars**}
\typeout{}
\def\draftdate{\today}
\def\mua{\marginpar[\boldmath\hfil$\uparrow$]%
                   {\boldmath$\uparrow$\hfil}%
                    \typeout{marginpar: $\uparrow$}\ignorespaces}
\def\mda{\marginpar[\boldmath\hfil$\downarrow$]%
                   {\boldmath$\downarrow$\hfil}%
                    \typeout{marginpar: $\downarrow$}\ignorespaces}
\def\mla{\marginpar[\boldmath\hfil$\rightarrow$]%
                   {\boldmath$\leftarrow $\hfil}%
                    \typeout{marginpar: $\leftrightarrow$}\ignorespaces}
\def\Mua{\marginpar[\boldmath\hfil$\Uparrow$]%
                   {\boldmath$\Uparrow$\hfil}%
                    \typeout{marginpar: $\Uparrow$}\ignorespaces}
\def\Mda{\marginpar[\boldmath\hfil$\Downarrow$]%
                   {\boldmath$\Downarrow$\hfil}%
                    \typeout{marginpar: $\Downarrow$}\ignorespaces}
\def\Mla{\marginpar[\boldmath\hfil$\Rightarrow$]%
                   {\boldmath$\Leftarrow $\hfil}%
                    \typeout{marginpar: $\Leftrightarrow$}\ignorespaces}
\overfullrule 5pt
\oddsidemargin -15mm
\marginparwidth 29mm
}

\makeatletter

\def\eqnarray{\stepcounter{equation}\let\@currentlabel=\theequation
\global\@eqnswtrue
\global\@eqcnt\z@\tabskip\@centering\let\\=\@eqncr
$$\halign to \displaywidth\bgroup\hskip\@centering
  $\displaystyle\tabskip\z@{##}$\@eqnsel&\global\@eqcnt\@ne
  \hskip 2\arraycolsep \hfil${##}$\hfil
  &\global\@eqcnt\tw@ \hskip 2\arraycolsep $\displaystyle\tabskip\z@{##}$\hfil
   \tabskip\@centering&\llap{##}\tabskip\z@\cr}
\def\appendix{\par
 \setcounter{section}{0} \setcounter{subsection}{0}
 \def\thesection{\Alph{section}}}

\makeatother

\newcommand{\lsim}
{\;\raisebox{-.3em}{$\stackrel{\displaystyle <}{\sim}$}\;}
\newcommand{\gsim}
{\;\raisebox{-.3em}{$\stackrel{\displaystyle >}{\sim}$}\;}

\begin{document}
\thispagestyle{empty}
\def\thefootnote{\fnsymbol{footnote}}
\setcounter{footnote}{1}
\null
\hfill BI-TP 96/06 \\
\null
\hfill KA-TP-03-96\\
\null
\hfill hep-ph/9602436
\vskip .8cm
\begin{center}
{\Large \bf How reliably can the Higgs-Boson Mass be predicted \\[.5em]
from Electroweak Precision Data?}%
\footnote{Partially supported by the EC-network contract CHRX-CT94-0579
and the Bundesministerium f\"ur Bildung und Forschung, Bonn, Germany.}
\vskip 2.5em
{\large S.\ Dittmaier, D.\ Schildknecht 
}
\vskip .5em
{\it Fakult\"at f\"ur Physik, Universit\"at Bielefeld, D-33615
Bielefeld, Germany}
\\[2ex]
{\large G.\ Weiglein}
\vskip .5em
{\it Institut f\"ur Theoretische Physik, Universit\"at Karlsruhe,
D-76128 Karlsruhe, Germany}
\vskip 2em
\end{center} \par
\vskip 1.2cm
\begin{center}
Revised version \\
June 1996
\end{center}
\vskip 1.2cm
\vfil
{\bf Abstract} \par
{}From the LEP precision data and the measurement of the W-boson mass, 
upon excluding the observables $\Rb$, $\Rc$
in a combined fit of the top-quark mass, $\Mt$, and the Higgs-boson
mass, $\MH$, within the Standard Model,
we find the weak $1\si$ bound of $\MH\lsim 900\GeV$.
Stronger upper bounds on $\MH$, sometimes presented in the literature,
rely heavily on the inclusion of $\Rb$ in the data sample. 
Upon including $\Rb$,
the quality of the fit drastically decreases, 
and by carefully analyzing the dependence of the fit results on the 
set of experimental input data we conclude that 
these stronger bounds 
are not reliable. Moreover, 
the stronger bounds on $\MH$ are lost if the deviation between 
theory and experiment in $\Rb$ is ascribed to contributions of new
physics. 
Replacing $\swbar^2(\LEP)$ by the combined value
$\swbar^2(\LEP+\SLD)$ in the data sample
leads to a bound of $\MH\lsim 430\GeV$ at the $1\si$ level. 
The value of $\swbar^2(\SLD)$ taken alone, however, gives rise to fit
results for $\MH$ 
which are in conflict with $\MH\gsim 65.2\GeV$ from direct 
searches.

\par
\vskip 1cm
\null
\setcounter{page}{0}
\clearpage
\def\thefootnote{\arabic{footnote}}
\setcounter{footnote}{0}

The discovery \cite{mtexp} of the top quark and the direct
determination of its mass of $\Mt^{\exp} = 180 \pm 12 \GeV$ open the
possibility of improving the 
constraints on the mass of the Higgs boson, $\MH$, from the body of the
precision electroweak data at the \PZ-boson 
resonance~\cite{data8/95,SLD} and the experimental 
value of the \PW-boson mass, $\MW$~\cite{MW}. 
In this note we present our results for
$\Mt$ and $\MH$, obtained by performing fits to 
the precision data and $\MW$ within the 
Standard Model (SM). The dependence of the fits on
the experimental data on $\Rb$, $\Rc$, and $\swbar^2$
is investigated, and the effects of varying the
SM input parameters $\alpz$ and $\alpsz$ in the allowed range are discussed.
We also examine how the results of these fits are influenced if
one allows for non-standard \Zbb,$\Pc\bar\Pc$ vertices. 
Even though several 
papers on this subject have appeared recently~\cite{Pass,Ell,Pok}, 
additional investigations combined with 
comments on the interpretation of the results seem useful.

We start from a global fit to the available electroweak
precision data. The large value of 
$\chidof$, obtained in the fit  
to be given below, requires a detailed analysis of the impact
of different parts of the experimental data. Accordingly, we will 
subsequently analyze the data in several distinct steps. 
In a first step we concentrate on the
leptonic observables $\Gl$ and $\swbar^2(\LEP)$, 
and $\MW$
(the set of data to be referred to as ``leptonic sector''), and include 
the total Z-boson width, $\GT$, and the Z-boson width into hadrons, 
$\Gh$,
from the set of hadronic observables (referring to this set of
data as ``all data $\backslash\;\Rb,\Rc$''), thus ignoring in the
fits at this stage the partial \PZ-boson decays \Zbb,$\Pc\bar\Pc$.
In a second step we include the \Zbb\ decay mode and
determine $\Mt$ and $\MH$ in a 
fit again. 
In a third step we finally discuss how the results of the fits
change when the decay \Zcc\ is included, 
and we also investigate the effect of replacing
$\swbar^2(\LEP)$ by $\swbar^2(\SLD)$ 
and by the combined 
value of $\swbar^2(\LEP+\SLD)$ in the set of data. 
Within all steps we carry out two alternative fits, a first one
in which 
the Tevatron result of $\Mt^{\exp} = 180 \pm 12$ GeV is included in the
fit, and a second one in which $\Mt$ is treated as a 
free fit parameter.
The procedure adopted obviously
allows us to identify the dependence of the 
results for $\Mt$ and $\MH$ on the set of experimental data
used in the fits. The various steps are motivated by the
discrepancies~\cite{data8/95} between 
SM prediction and experiment observed in the 
\Zbb,$\Pc\bar\Pc$ decays 
and the difference between the LEP and SLD results for $\swbar^2$.
By including/excluding the experimental information on $\Mt$ we
furthermore investigate how strongly the fit results for $\Mt$ and $\MH$
are correlated.
In a final step we discuss fits in which non-standard
contributions are allowed.

{\doublerulesep 3pt
\btab
\bce
\begin{tabular}{|@{}c@{}||c|c|}
\hline
leptonic sector & \multicolumn{2}{c|}{hadronic sector} \\
\hline
\hline
$\Gl = 83.93 \pm 0.14 \MeV$ & 
$R = 20.788 \pm 0.032$ & $\GT = 2496.3 \pm 3.2 \MeV$ \\
\hline
$\swbar^2 (\LEP) = 0.23186 \pm 0.00034$ & 
$\si_{\mathrm h} = 41.488 \pm 0.078$ &
$\Gh = 1744.8  \pm 3.0 \MeV$ \\
\hline
$\swbar^2 (\SLD) = 0.23049 \pm 0.00050$ & 
$\Rb = 0.2219 \pm 0.0017$ &
$\Gb = 387.2 \pm 3.0 \MeV$ \\
\hline
$\swbar^2 (\LEP+\SLD) = 0.23143\pm0.00028$ &
$\Rc = 0.1543 \pm 0.0074$ &
$\Gc = 269 \pm 13 \MeV$ \\ \hline
$\MW = 80.26 \pm 0.16 \GeV$ & \multicolumn{2}{c|}{} \\
\hline \hline
input parameters & \multicolumn{2}{c|}{correlation matrices} \\
\hline \hline
\begin{array}[b]{c}
\MZ = 91.1884 \pm 0.0022 \GeV \\ \hline
\hspace{20pt} \GF = 1.16639 (2) \cdot 10^{-5} \GeV^{-2} \hspace{20pt} \\ \hline
\alpz^{-1} = 128.89 \pm 0.09 \\ \hline
\alpsz = 0.123\pm0.006 \\ \hline
\Mb = 4.5\GeV \hspace{1pt} \\ \hline
\Mt = 180 \pm 12 \GeV\\
\end{array}
& \multicolumn{2}{c|}{
\begin{array}[b]{|c||c|c|c|c|}
\hline
& \si_{\mathrm h} & R & \GT \\ \hline\hline
\si_{\mathrm h} & \phantom{-}1.00 & \phantom{-}0.15 & -0.12 \\ \hline
R               & \phantom{-}0.15 & \phantom{-}1.00 & -0.01 \\ \hline
\GT             & -0.12 & -0.01 & \phantom{-}1.00 \\ \hline
\earr
\qquad
\begin{array}[b]{|c||c|c|c|}
\hline
& \Rb & \Rc \\ \hline\hline
\Rb  & \phantom{-}1.00 & -0.34 \\ \hline
\Rc  & -0.34 & \phantom{-}1.00 \\ \hline
\earr} \\ \hline
\end{tabular}
\ece
\caption[]{
The precision data used in the fits, consisting of the LEP data
\cite{data8/95}, the SLD value \cite{SLD} for $\swbar^2$, and the world 
average \cite{MW} for $\MW$.
The partial widths $\Gl$, $\Gh$, $\Gb$, and $\Gc$ are obtained from the
observables $R = \Gh/\Gl$, $\si_{\mathrm h} =
(12\pi\Gl\Gh)/(\MZ^2\Ga^2_{\mathrm T})$, 
$\Rb =  \Gb/\Gh$,
$\Rc =  \Gc/\Gh$, and  $\GT$ using the given correlation
matrices. The data in the upper left-hand column (using $\swbar^2 (\LEP)$
if not otherwise specified) will be referred to as
``leptonic sector'' in the fits. Inclusion of the data in the upper
right-hand column will be referred to as fitting ``all data''.
The theoretical predictions are based on the input 
parameters~\protect\cite{mtexp,data8/95,bu95} 
given in
the lower left-hand column of the table.}
\label{tab:data}
\etab
}

The set of experimental data 
is listed in \refta{tab:data}. 
A subset of the data in \refta{tab:data} is referred to as 
``input parameters''.
This is motivated by the high experimental accuracy of some of the 
quantities (namely $\GF$ and $\MZ$) and by the non-electroweak origin of 
others $(\alpz, \alpsz$). 
The listed ``input parameters'' represent the 
commonly 
used input for
theoretical predictions in the on-shell renormalization scheme.
The experimental error
of $\GF$ is entirely negligible with respect to the determination of 
$\Mt$ and $\MH$. This is also true for $\MZ$. 
For completeness, this was 
explicitly verified by treating $\MZ$ as additional fit parameter.
If not otherwise indicated,
the parameter $\alpz$ will be treated as fit parameter 
employing the constraint of \refta{tab:data}. 
Finally, we note that the value of $\alpsz$ given in \refta{tab:data}
is the result from the LEP event shape analysis~\cite{data8/95}.
Due to the fact that this value disagrees with results from different
experiments (e.g. deep-inelastic scattering) and
lattice calculations, $\alpsz$ 
will either be treated as free fit parameter or the influence of varying
$\alpsz$ as input parameter will be studied separately. For the input parameter 
$\Mb$, the mass of the $\Pb$ quark,  
the value of $\Mb$
from \refta{tab:data} will be inserted. A 
detailed analysis reveals that the results for $\Mt$ and $\MH$ are 
independent of the precise value of $\Mb$ for any
reasonable changes of $\Mb$. Otherwise, the notation in \refta{tab:data} is 
standard. The partial Z-boson width into a lepton and an anti-lepton, assuming
universality, is denoted by $\Gl$. 
The partial widths for \Zbb\ and \Zcc\ are given by $\Gb$ and $\Gc$.
Finally, the effective electroweak angle, $\swbar^2$, in \refta{tab:data} is 
defined by the effective vector and axial vector couplings 
($g_{\mathrm{V,l}}$ and $g_{\mathrm{A,l}}$, respectively) of the Z~boson to 
leptons at the Z resonance, 
$\swbar^2 \equiv \sin^2 \theta^{{\mathrm lept}}_{{\mathrm eff}} 
\equiv (1-g_{\mathrm{V,l}}/g_{\mathrm{A,l}})/4$. It is accordingly
extracted from the asymmetry measurements at LEP~\cite{data8/95}
and SLD~\cite{SLD}.

The theoretical 
SM results at the one-loop level, taking into
account leading two-loop contributions, are taken from 
\citeres{zph1,zph2}%
\footnote{We have supplemented the analytical results given in 
\citeres{zph1,zph2}
by the ${\cal O}(G_\mu\Mt^2\alps^2)$ corrections \cite{3looprho} to
the $\rho$-parameter.}. 
Therefore, we provide an analysis which is completely independent 
of results presented by other authors \cite{Pass,Ell,Pok}.

We obtain for the global fit to the complete set of 
data listed in \refta{tab:data}
\beq
\begin{array}[b]{lll}
\Mt=167^{+11}_{-9}\GeV, & \MH=81^{+144}_{-52}\GeV, & \\
\alpz^{-1}=128.90\pm0.09, \quad &  \alpsz=0.121\pm0.004, \quad &
\chidof=17/9,
\end{array}
\label{eq:globfit}
\eeq
where the combined value $\swbar^2(\LEP + \SLD)$ has been used.
In this fit $\alpsz$ has been used as a free fit parameter, while the
experimental constraints on $\alpz$ and $\Mt$ have been included. 
The result \refeq{eq:globfit} is in good agreement with the
corresponding results given in \citeres{Ell,Pok}.%
\footnote{
In \citere{Ell} also the available low-energy data were
included in the analysis, which shows that the effect of these data on
the results of the SM fits is rather small.}
While the low central value and the rather tight $1 \sigma$ bounds obtained for
$\MH$ in this fit seem to indicate evidence for a light Higgs-boson
mass, the high value of $\chidof =  17/9$ gives rise to the question 
of how reliable this bound actually is.

In order to investigate the dependence of the fit results on
inclusion/exclusion of different parts of the experimental data, we now
turn to an analysis in distinct steps as outlined above. The results
for the corresponding fits of the parameters $(\Mt, \MH, \alpz, \alpsz)$
within the SM are presented in \refta{tab:globfit} and in the
$(\MH,\Delta\chi^2)$-plots of \reffi{fig:Dchi}. In these fits $\alpsz$
is treated as a free fit parameter while $\alpz$ is fitted 
including
the experimental constraint from \refta{tab:data}. 
Note that the values obtained for $\alpsz$ in these fits practically
coincide with the value of \refta{tab:data} which is deduced
by the entirely different method of an event-shape (jet production) 
analysis. 
As mentioned above, $\Mt$ is treated in two different ways in the fits.
Treating $\Mt$ as a free fit parameter allows
to compare its fit result with its actual experimental value, while
using this information in the fit from the start leads to a certain
``compromise'' result which might be more difficult to interpret.

In \reffi{fig:mtmhfit} we furthermore investigate the dependence of the
fit results on variations in $\alpz$ and $\alpsz$. To this end these
parameters are not fitted but kept as fixed values which are varied
within the $1\sigma$ bounds of their experimental values given in
\refta{tab:data}. The top-quark
mass is treated as a free fit parameter in this figure. In the last row
of \reffi{fig:mtmhfit} the effect of replacing $\swbar^2 (\LEP)$
by $\swbar^2 (\LEP + \SLD)$ and by $\swbar^2 (\SLD)$ is studied. 

\btab
Table \ref{tab:globfit}a: \\[.2cm]
\begin{tabular}{|l||l|l|c|c|}
\hline
\rlap{using \boldmath{$\swbar^2(\LEP)$}}%
\phantom{using \boldmath{$\swbar^2(\LEP+\SLD)$}}
& \multicolumn{1}{c|}{$\Mt/\GeV$} & \multicolumn{1}{c|}{$\MH/\GeV$} &
$\alpsz$ 
& $\chidof$ \\
\hline \hline
leptonic sector ${}+ \Mt^{\exp}$ &
$179^{+12}_{-11}$ &
$353^{+540}_{-224}$ & 0.123 (fixed) 
& $0.2/5$ \\
\hline
leptonic sector &
$174^{+37}_{-19}$ &
$248^{+\nolimit}_{-194}$ & 0.123 (fixed) 
& $0.2/4$ \\
\hline \hline
all data ${}+ \Mt^{\exp}$ $\backslash$ $\Rb,\Rc$ &
$179^{+12}_{-12}$ & $356^{+543}_{-227}$ &
$0.124^{+0.004}_{-0.004}$ 
& 0.7/7 \\
\hline
all data \phantom{${}+ \Mt^{\exp}$} $\backslash$ $\Rb,\Rc$ &
$167^{+45}_{-20}$ &
$163^{+\nolimit}_{-126}$ &
$0.123^{+0.007}_{-0.004}$ 
& 0.6/6 \\
\hline \hline
all data ${}+ \Mt^{\exp}$ $\backslash$ $\Rb$ &
$178^{+12}_{-12}$ & $343^{+523}_{-219}$ &
$0.124^{+0.004}_{-0.004}$ 
& 6.6/8 \\
\hline
all data \phantom{${}+ \Mt^{\exp}$} $\backslash$ $\Rb$ &
$164^{+40}_{-18}$ & $133^{+\nolimit}_{-97}$ &
$0.123^{+0.006}_{-0.004}$ 
& 6.4/7 \\
\hline \hline
all data ${}+ \Mt^{\exp}$ $\backslash$ $\Rc$  &
$169^{+11}_{-11}$ & $186^{+277}_{-119}$ &
$0.123^{+0.004}_{-0.004}$ 
& $15/8$ \\
\hline
all data \phantom{${}+ \Mt^{\exp}$} $\backslash$ $\Rc$  &
$148^{+14}_{-12}$ & $54^{+93}_{-30}$ &
$0.122^{+0.004}_{-0.004}$ 
& $12/7$ \\
\hline \hline
all data ${}+ \Mt^{\exp}$ &
$170^{+11}_{-11}$ & $197^{+291}_{-126}$ &
$0.123^{+0.004}_{-0.004}$ 
& $16/9$ \\
\hline
all data &
$149^{+15}_{-12}$ & $57^{+104}_{-32}$ &
$0.122^{+0.004}_{-0.004}$ 
& $14/8$ \\
\hline
\end{tabular}
\\[.3cm]
Table \ref{tab:globfit}b: \\[.2cm]
\begin{tabular}{|l||l|l|c|c|c|}
\hline
using \boldmath{$\swbar^2(\LEP+\SLD)$}
& \multicolumn{1}{c|}{$\Mt/\GeV$} & \multicolumn{1}{c|}{$\MH/\GeV$} &
$\alpsz$ 
& $\chidof$ \\
\hline \hline
leptonic sector ${}+ \Mt^{\exp}$ &
$175^{+12}_{-11}$ &
$152^{+282}_{-106}$ & 0.123 (fixed) 
& $1.0/5$ \\
\hline
leptonic sector &
$165^{+18}_{-10}$ &
$64^{+223}_{-37}$ & 0.123 (fixed) 
& $0.3/4$ \\
\hline \hline
all data ${}+ \Mt^{\exp}$ $\backslash$ $\Rb,\Rc$ &
$176^{+12}_{-12}$ & $154^{+273}_{-108}$ &
$0.122^{+0.004}_{-0.004}$ 
& 1.5/7 \\
\hline
all data \phantom{${}+ \Mt^{\exp}$} $\backslash$ $\Rb,\Rc$ &
$161^{+21}_{-12}$ &
$51^{+214}_{-31}$ &
$0.121^{+0.007}_{-0.004}$ 
& 0.7/6 \\
\hline \hline
all data ${}+ \Mt^{\exp}$ $\backslash$ $\Rb$ &
$175^{+12}_{-11}$ & $148^{+263}_{-103}$ &
$0.122^{+0.004}_{-0.004}$ 
& 7.3/8 \\
\hline
all data \phantom{${}+ \Mt^{\exp}$} $\backslash$ $\Rb$ &
$160^{+19}_{-12}$ & $49^{+174}_{-29}$ &
$0.121^{+0.004}_{-0.004}$ 
& 6.5/7 \\
\hline \hline
all data ${}+ \Mt^{\exp}$ $\backslash$ $\Rc$  &
$167^{+11}_{-9}$ & $76^{+136}_{-49}$ &
$0.121^{+0.004}_{-0.004}$ 
& $15/8$ \\
\hline
all data \phantom{${}+ \Mt^{\exp}$} $\backslash$ $\Rc$  &
$152^{+11}_{-11}$ & $34^{+46}_{-17}$ &
$0.122^{+0.004}_{-0.004}$ 
& $12/7$ \\
\hline \hline
all data ${}+ \Mt^{\exp}$ &
$167^{+11}_{-9}$ & $81^{+144}_{-52}$ &
$0.121^{+0.004}_{-0.004}$ 
& $17/9$ \\
\hline
all data &
$153^{+11}_{-11}$ & $35^{+50}_{-18}$ &
$0.121^{+0.004}_{-0.004}$ 
& $14/8$ \\
\hline
\end{tabular}
\caption[]{
The results obtained in $(\Mt,\MH,\alpz,\alpsz)$
fits to different sets of experimental data, as indicated (see text). 
The results in \refta{tab:globfit}a are based on $\swbar^2(\LEP)$, while the 
results in \refta{tab:globfit}b are based on $\swbar^2(\LEP+\SLD)$. For each 
set of experimental data, the fit results given in the lower row are 
obtained by treating $\Mt$ as a free fit parameter, while the results in 
the upper row include the constraint $\Mt^{\exp}=180\pm12\GeV$.
Note that the fit results on $\alpz$ are not explicitly stated,
because they range between $\alpz^{-1}=128.89\pm0.09$ and
$\alpz^{-1}=128.91\pm0.09$ for all cases, thus reproducing the input
value from \refta{tab:data}. 
}
\label{tab:globfit}
\etab

We first of all concentrate on the results of {\it the first step of our
analysis}, namely the fits in \refta{tab:globfit}a and \reffi{fig:mtmhfit} 
based on the data sets $\swbar^2(\LEP)$, $\Gl$ and $\MW$ 
(``leptonic sector'') 
and $\swbar^2(\LEP)$, $\Gl$, $\MW$, $\GT$, $\Gh$ 
(``all data $\backslash\;\Rb,\Rc$''). Both fits yield an excellent 
$\chidof<1$, independently of whether $\alpz$ and $\alpsz$ are fitted or
whether they are taken as fixed input parameters that are varied within
one standard deviation according to \refta{tab:data}.
Figure~\ref{fig:mtmhfit}
shows that the results of the fits
are strongly affected by
variations of $\alpz^{-1}$. For instance, lowering 
$\alpz^{-1} = 128.89$ by one standard deviation to $\alpz^{-1} = 128.80$ 
also lowers the central value of $\MH$ by approximately one standard
deviation. 
Varying $\alpsz$ from $\alpsz=0.123$ to $\alpsz=0.117$ shifts the
upper $1\sigma$ limit
of $\MH$ from $\sim 1\TeV$ to $265\GeV$ in the fit 
in which $\Gh$ and $\GT$ are included 
(second column of \reffi{fig:mtmhfit}).
We also note the somewhat low values of the top-quark mass of 
$\Mt=157\GeV$ and $\Mt=162\GeV$ obtained for the lower values of 
$\alpz^{-1}$ and $\alpsz$, respectively, 
which are below the $1\sigma$ lower limit of $\Mt = 168\GeV$
from the direct measurement of $\Mt$. 
The fit results in the leptonic sector are stable 
under variation in the strong coupling constant, $\alpsz$, 
since $\alpsz$ only enters at the two-loop level.
Altogether, we thus conclude that 
varying $\alpz^{-1}$ and $\alpsz$ within the $1\sigma$ bounds given in
\refta{tab:data} leads
to a considerable effect concerning the boundaries in the $(\Mt,\MH)$
plane. 
Consequently, concerning the range of $\MH$ allowed by the
results from the leptonic sector (with $\swbar^2(\LEP)$) 
and $\GT, \Gh$, 
i.e.\ by fitting the SM only to those data that agree with the
theoretical predictions,
according to the foregoing discussion of 
\refta{tab:globfit}a and \reffi{fig:mtmhfit},
it seems hardly possible to deduce stronger limits than 
$\MH\lsim 900\GeV$
at the $1\sigma$ level, even upon taking into account the constraint of 
$\Mt^{\exp}=180\pm12\GeV$ from the direct observation of the top quark.

We turn to the {\it second step of our analysis} and include $\Rb$ in the 
fit, which is thus based on the leptonic sector in conjunction with $\GT$, 
$\Gh$ and $\Rb$.
According to \refta{tab:globfit}a and \reffi{fig:mtmhfit}, taking into 
account the data for the \Zbb\ partial width leads to an increase of $\chidof$
by about an order of magnitude.
Comparing the 
third column in \reffi{fig:mtmhfit} with the first and second columns, one 
observes a considerable shrinkage of the $1\sigma$ 
regions in the $(\Mt,\MH)$ plane and a 
drastic shift towards lower values of $\Mt$ and $\MH$. The
sensitivity against variations of $\alpz^{-1}$ and $\alpsz$
is considerably weaker in this sample of data. 
The central fit-value for the top-quark mass of $\Mt=148^{+14}_{-12}\GeV$ 
(where the experimental constraint on $\Mt$ has not been taken into 
account in the fit) 
is significantly below the central value of the direct measurement 
of $\Mt^{\exp}=180\pm12\GeV$, and the central value obtained for the 
Higgs-boson mass, $\MH=54^{+93}_{-30}\GeV$,
lies in the vicinity of the 
experimental lower bound $\MH>65.2\GeV$. 

The large increase of $\chidof$ 
when including $\Rb$ in the fit 
signals the large discrepancy between theory and experiment in this fit.
In particular, when evaluating $\Rb$ for the best-fit values of
$(\Mt,\MH)=(148^{+14}_{-12}\GeV,54^{+93}_{-30}\GeV)$,
the resulting ($\MH$-insensitive) theoretical prediction,
$\Rb^\SM = 0.2164^{-0.0005}_{+0.0004}$
(with the errors indicating the changes by varying $\Mt$ 
within the $1\si$ limits),
still lies more than $3\sigma$ below 
the experimental value of $\Rb = 0.2219 \pm 0.0017$.

\btab
\bce
\begin{tabular}{|c||c|c|c|c|}
\hline
$\Mt/\GeV$ fixed & 144 & 168 & 180 & 192\\ \hline
& \multicolumn{4}{c|}{$\MH/\GeV$ ($\chidof$)} \\ 
\hline \hline
leptonic sector & $47^{+30}_{-18}$ (4.2/3) & $160^{+104}_{-69}$ (0.2/3) & 
$362^{+206}_{-136}$ (0.2/3) & $792^{+444}_{-280}$ (0.4/3) \\
\hline
all data $\backslash$ $\Rb$, $\Rc$ &
$44^{+34}_{-20}$ (2.1/5) &
$172^{+110}_{-73}$ (0.6/5) & $349^{+196}_{-131}$ (0.8/5) &
$682^{+368}_{-239}$ (1.3/5) \\
\hline
all data $\backslash$ $\Rb$ &
$44^{+34}_{-20}$ (7.7/6) &
$174^{+111}_{-73}$ (6.4/6) & $353^{+199}_{-133}$ (6.7/6) &
$689^{+375}_{-242}$ (7.3/6) \\
\hline
all data $\backslash$ $\Rc$ &
$45^{+35}_{-21}$ (12/6) & $176^{+112}_{-74}$ (14/6) & 
$355^{+199}_{-133}$ (16/6) &
$685^{+368}_{-239}$ (19/6) \\
\hline
all data & $45^{+35}_{-21}$ (14/7) & 
$176^{+112}_{-74}$ (15/7) & 
$355^{+199}_{-133}$ (17/7) & $686^{+369}_{-240}$ (20/7)\\
\hline
\end{tabular}
\ece
\caption{Fits of $\MH$ to various sets of experimental data in the 
SM for fixed values of $\Mt$.
In all fits $\alpz^{-1} = 128.89$ and $\alpsz = 0.123$ are kept fixed, and
the LEP value of $\swbar^2$ is used in the input data.}
\label{tab:mtmhfitfix}
\etab

In connection with the low central fit
value of $\Mt=148\GeV$, it is illuminating 
to consider the results of single-parameter $\MH$ fits, where
$\Mt$ is kept fixed at certain (assumed) values. 
In \refta{tab:mtmhfitfix}, 
again for the previously selected sets of data, 
results of single-parameter $\MH$ fits are shown.
The known strong $(\Mt,\MH)$ correlation in SM fits
leads to a remarkable stability of
the resulting fit values for $\MH$. Once $\Mt$ is fixed, 
there is almost no dependence of the fit value for $\MH$ on which set of
input data is actually used in the fit. In particular, whenever a low
value of $\Mt$ is chosen, one obtains a low value for $\MH$,
independently of whether $\Rb$ is included in the fit or not.%
\footnote{Conversely, if $\MH$ is fixed, the values of $\Mt$ obtained in
the fit are fairly stable, independently of whether $\Rb$ or $\Rc$ are
included in the fit or not. This is consistent with the results of
\citere{data8/95},
where $\MH=300\GeV$ is kept fixed when fitting $\Mt$.} 
Since the ($\MH$-insensitive) SM prediction for $\Rb$
increases with decreasing $\Mt$, 
in the combined fit of $(\Mt,\MH)$
the inclusion of $\Rb$ lowers 
the fit value of $\Mt$, and via the $(\Mt,\MH)$ correlation also the
value of $\MH$. As discussed above, the result of $\Mt=148\GeV$ is
nothing but a kind of compromise, as it still leads to a $3\si$
discrepancy between theory and experiment in $\Rb$.
Moreover, this result for $\Mt$ is disfavored by
the Tevatron result of $\Mt^{\exp}=180\pm 12\GeV$. 

While the problematic features of the fits where $\Rb$ is included 
are easy to see in the case where $\Mt$ is used as a free fit parameter,
they are somewhat hidden in the fits where the experimental information
on $\Mt$ is used. It partially compensates the tendency of the fits
towards low values of $\Mt$ and leads to the more moderate looking result
of $\Mt = 169 \pm 11 \GeV$ and $\MH = 186^{+277}_{-119} \GeV$. In view of
the foregoing discussion, however, the result for $\MH$ obtained in this
way appears to be rather questionable.
 
In summary, the large value of $\chidof$ and the
low fit value for $\Mt$ (when $\Mt$ is treated as free fit parameter)
that is at variance with the Tevatron result, lead to
the conclusion that the low value 
and tight bound obtained for $\MH$ when
including the data for $\Rb$ does not seem 
reliable.  It is an artifact of
the procedure of describing the ``non-standard''
value of $\Rb$ by the unmodified SM in conjunction with
the $(\Mt,\MH)$ correlation.
This conclusion is strengthened by the fact that a
simple phenomenological modification of the \Zbb\ vertex,
to be discussed below, leads to values of $\Mt$ 
compatible with the Tevatron result
and removes the stringent upper bounds on $\MH$. 

We turn to the third step of our analysis and consider
the {\it impact of the observable $\Rc$}.
As can be seen in \refta{tab:globfit}, the results for $\Mt$ and
$\MH$ are hardly affected by including $\Rc$.
This is a consequence of the fact that the contribution of 
$\Rc$ to $\chi^2$ depends only very weakly on $\Mt$ and $\MH$, because 
the experimental error for $\Rc$ is much larger than the change in
the SM prediction for $\Rc$ induced by varying $\Mt$ and $\MH$.
Similarly to the case of $\Rb$, including $\Rc$ in the set of data (and
omitting $\Rb$) leads to an enhanced value of $\chidof$
and to a tendency towards lower values of $\MH$.

So far the analysis has been based on the LEP experimental value of 
$\swbar^2(\LEP)$. 
Table~\ref{tab:globfit}b and the last row of
\reffi{fig:mtmhfit} show the
{\it effect of replacing $\swbar^2(\LEP)$ by $\swbar^2(\LEP + \SLD)$}.
In \reffi{fig:mtmhfit} also contours are shown that are based on
taking $\swbar^2(\SLD)$ alone. The
change in the allowed $(\Mt,\MH)$ plane occurring as a consequence of 
these replacements is very strong.
For the ``leptonic sector'' and ``all data $\backslash\;\Rb,\Rc$''
the fit value of $\Mt\simeq 170\GeV$ 
(where $\Mt^{\exp}$ has not been included in the fit)
is consistent with
the value from the direct measurements, $\Mt^{\exp}=180\pm12\GeV$, while
the values for $\MH$ resulting from using $\swbar^2(\SLD)$ now have 
decreased to 
$\MH = 18^{+28}_{-9} \GeV$ 
and $\MH = 16^{+27}_{-9} \GeV$,
respectively. The fit to 
``all data'' 
using $\swbar^2(\SLD)$ yields 
$\Mt=161^{+10}_{-11}\GeV$ and a similarly low value for 
$\MH$, namely
$\MH=18^{+24}_{-9}\GeV$.
Accordingly, using the SLD value for $\swbar^2$ leads to very low
fit results for $\MH$, independently of whether $\Rb$ and $\Rc$
are included in the data set or not, 
and of whether use is made
of the experimental information on $\Mt$.
Comparing these results to the lower bound from the direct Higgs-boson
search, $\MH>65.2\GeV$~\cite{MH}, 
one arrives at a serious conflict between the unmodified SM and
experiment.
The discrepancy is weakened if the combined value of
$\swbar^2(\LEP+\SLD)$ from \refta{tab:globfit}b is used. In this case
one obtains
$\Mt=153 \pm 11\GeV$ and $\MH=35^{+50}_{-18}\GeV$ for ``all data''
(where again $\Mt^{\exp}$ has not been included).
A resolution of the LEP--SLD discrepancy on $\swbar^2$ is obviously one
of the most important tasks with respect to the issue of $\MH$ bounds
via radiative corrections.

As a summary of the present situation concerning $\MH$, in 
\reffi{fig:Dchi} we present the result of selected 
$(\Mt,\MH,\alpz,\alpsz)$ fits according to \refta{tab:globfit} in a
$(\MH,\De\chi^2)$ plot. The quantitative influence on the fit value of
$\MH$ resulting from inclusion of $\Mt^{\exp}=180\pm12\GeV$ can be seen
to agree with the qualitative expectations from \reffi{fig:mtmhfit}.
Other features of the results for $\MH$ previously read off from
\reffi{fig:mtmhfit}, such as the correlation between $\MH$ and the input
for $\swbar^2$, or the effect of ignoring the experimental results for
$\Rb$, $\Rc$ can obviously also be seen in \reffi{fig:Dchi}. The plots
in \reffi{fig:Dchi} clearly illustrate the difficulty of establishing a
unique bound on $\MH$. The most reliable bound, from ``all data
$\backslash\,\Rb,\Rc$'', but including $\Mt^{\exp}$ yields 
$\MH\lsim430\GeV$ based on $\swbar^2(\LEP +\SLD)$, and $\MH\lsim900\GeV$
based on $\swbar^2(\LEP)$.

In order to accommodate the experimental result for $\Rb$, we now
allow for a {\it modification of the \Zbb\ vertex}
by a parameter $\De\yb$, as introduced in \citere{zph2}.%
\footnote{The parameter $\De\yb$ is related to the parameter
$\eps_\Pb$ introduced in \citere{Alteps} via
$\De\yb=-2\eps_\Pb-0.2\times10^{-3}$ (see \citere{zph2}).}
The possible origin of this modification of the 
SM predictions is left open for the time being, but 
in particular it includes the impact of new particles 
in conjunction with loop corrections at the \Zbb\ vertex.
We allow for values of $\alpsz$ different~\cite{Shif&Alt}
from the LEP value of $\alpsz = 0.123$, in order to compensate
for the enhanced theoretical value of the total hadronic \PZ-boson
width, $\Gh$, resulting from the enlarged theoretical value of $\Gb$ 
which is adjusted to be in agreement with experiment.
Deviations of $\De\yb$ from its ($\Mt$-dependent) SM value
$\De\yb^{\SM}$~\cite{zph2} 
lead to an extra contribution $X_\Pb$ \cite{data8/95} 
in the prediction for $\Gb$,
\beqar
X_\Pb = \Gb - \Gb^{\SM} &=& 
\frac{\alpz\MZ}{24s_0^2c_0^2}(2s_0^2-3)
\,R_\QED\,R_\QCD\,\left(\De\yb-\De\yb^{\SM}\right) \nn\\
&=& -0.421{\GeV}\,\times\,R_\QCD\,\left(\De\yb-\De\yb^{\SM}\right),
\eeqar
where
$s^2_0c^2_0=s^2_0(1-s^2_0)=\pi\alpz/\sqrt{2}\GF\MZ^2$,
$R_\QED=1+\alpha/12\pi$,
and $R_\QCD=1+\alpsz/\pi+1.41(\alpsz/\pi)^2-12.8(\alpsz/\pi)^3$ 
according to \citere{PDG}.

\btab
Table \ref{tab:dyb}a (``all data ${}+ \Mt^{\exp}$ $\backslash\;\Rc$''): 
\\[.2cm]
\begin{tabular}{|c||c|c|c|c|}
\hline
$\alpsz$ & $\Mt/\GeV$ & $\MH/\GeV$ & $\De y_\Pb/10^{-3}$ & 
$\chidof$\\
\hline \hline
$0.123$ fixed & $179^{+11}_{-11}$ &
$582^{+\nolimit}_{-324}$ & $3.9^{+4.6}_{-4.6}$ & 
$11/8$ \\ 
\hline
$0.110$ fixed & $179^{+11}_{-11}$ &
$523^{+\nolimit}_{-302}$ & $-8.8^{+4.6}_{-4.6}$ &
$3.4/8$ \\ 
\hline
$0.100$ fixed & $179^{+12}_{-11}$ &
$472^{+\nolimit}_{-284}$ & $-18.6^{+4.6}_{-4.6}$ & 
$0.9/8$ \\ 
\hline \hline
$0.098 \pm 0.008$ fitted & $179^{+12}_{-12}$ & $459^{+\nolimit}_{-281}$ & 
$-20.9^{+8.9}_{-8.9}$ & 
$0.9/8$ \\
\hline
\end{tabular}
\\[.3cm]
Table \ref{tab:dyb}b (``all data $\backslash\;\Rc$''):  \\[.2cm]
\begin{tabular}{|c||c|c|c|c|c|}
\hline
$\alpsz$ & $\Mt/\GeV$ & $\MH/\GeV$ & $\De y_\Pb/10^{-3}$ & 
$\chidof$\\
\hline \hline
$0.123$ fixed & $173^{+28}_{-22}$ &
$414^{+\nolimit}_{-294}$ & $3.8^{+4.6}_{-4.6}$ & 
$11/7$ \\ 
\hline
$0.110$ fixed & $174^{+32}_{-23}$ &
$375^{+\nolimit}_{-284}$ & $-8.8^{+4.6}_{-4.6}$ &
$3.3/7$ \\ 
\hline
$0.100$ fixed & $172^{+37}_{-24}$ &
$300^{+\nolimit}_{-236}$ & $-18.6^{+4.7}_{-4.6}$ & 
$0.9/7$ \\ 
\hline \hline
$0.098 \pm 0.008$ fitted & $171^{+39}_{-24}$ & 
$269^{+\nolimit}_{-216}$ & $-21.0^{+8.9}_{-9.0}$ &
$0.8/7$ \\
\hline
\end{tabular}
\caption[xxx]{
The results of four-parameter and five-parameter fits to 
``all data $\backslash\;\Rc$'' with $\Mt^{\exp}$ in-/excluded.
In the four-parameter
$(\Mt,\MH,\De\yb,\alpz)$ fits $\alpsz$ is kept fixed as indicated (first
three rows), while in the five-parameter $(\Mt,\MH,\De\yb,\alpz,\alpsz)$
fits (last row) $\alpsz$ is treated as a free fit parameter.
The fit results for $\alpz$ are again omitted, since they merely vary
between $\alpz^{-1}=128.90 \pm 0.09$ and $\alpz^{-1}=128.92 \pm 0.09$.
}
\label{tab:dyb}
\etab
In \refta{tab:dyb} we present 
our results for four-parameter $(\Mt, \MH, \De\yb, \alpz)$ fits with
fixed values of $\alpsz$, 
as well as the results of 
five-parameter $(\Mt, \MH, \De\yb, \alpz, \alpsz)$ fits. 
Table~\ref{tab:dyb}a is based on the data set 
``all data ${}+ \Mt^{\exp}$ $\backslash\;\Rc$''
(experimental information on $\Mt$ included),
while in \refta{tab:dyb}b we have used ``all data $\backslash\;\Rc$''
($\Mt$ treated as a free fit parameter).
The conclusion from \refta{tab:dyb}
is simple: once one allows for a modification of
$\Rb$ by the parameter $\De\yb$, the bounds on $\MH$ obtained by
fitting within the unmodified SM are lost. The quality of the fit 
is improved considerably, if one allows for a value of $\alpsz$ 
substantially below the LEP result from the event shape 
measurement~\cite{data8/95} of $\alpsz = 0.123 \pm 0.006$.
Fitting also $\alpsz$ leads to the extremely low best-fit value of 
$\alpsz = 0.098 \pm 0.008$, 
which is even lower than the extrapolated value of $\alpsz$ from
low-energy deep inelastic scattering data, 
$\alpsz = 0.112 \pm 0.004$~\cite{PDG}, and the value obtained from lattice 
QCD calculations, $\alpsz = 0.115 \pm 0.003$~\cite{PDG}.
The values of $\Mt$ in \refta{tab:dyb}b roughly coincide with the ones
obtained in the SM fits to the ``leptonic sector'' given in
\refta{tab:globfit}a. For low $\alpsz$ also the $\MH$ bounds in
\refta{tab:dyb}b are similar to the results of the SM fit obtained for
the ``leptonic sector'' (\refta{tab:globfit}a).

As in the previous case of the pure SM fits, 
the results do not change qualitatively when
$\Rc$ is included in the data set. 
In the $(\Mt,\MH,\De\yb,\alpz,\alpsz)$ fit to ``all data ${}+ \Mt^{\exp}$''
we obtain
\beq
\begin{array}[b]{rcl}
\Mt &=& 179 \pm 12\GeV, \qquad
\MH = 440^{+\nolimit}_{-270}\GeV, \qquad
\alpz^{-1} = 128.90 \pm 0.09,\\[.3em]
\alpsz &=& 0.102\pm0.008, \qquad
\De\yb = (\,-15.2^{+8.5}_{-8.6}\,)\times 10^{-3}, 
\qquad
\chidof = 5.8/9.
\end{array}
\label{eq:mtmhdybalpsz}
\eeq
This is in good agreement with the results presented in
\citere{data8/95} for a fit of $(\Mt,\alps,X_\Pb)$ for $\MH$
fixed at $\MH=300\GeV$.
The increased value of $\chidof$ in \refeq{eq:mtmhdybalpsz} relative to
the corresponding value in \refta{tab:dyb}
is of course a consequence of the
$2.5\sigma$ discrepancy \cite{data8/95} in $\Rc$. 
However, it does not seem to be meaningful to introduce an additional 
non-standard parameter $\De\yc$ in order to accommodate the $\Rc$ 
discrepancy. On the one hand, a modification of the
\Zcc\ vertex is much less motivated than in the case of the \Zbb\
vertex (see e.g.\ the discussion in \citere{zph2}); 
on the other hand, a fit in which 
a non-standard $\De\yc$ is allowed yields the absurd
value $\alpsz=0.19\pm0.04$, which was 
also obtained e.g.\ in \citeres{data8/95,Pok,PDG}.

\btab
\bce
\begin{tabular}[t]{|r||r|r||r|r|}
\hline
\multicolumn{1}{|c||}{$\alpsz$}
& \multicolumn{1}{c|}{$0.123$} & \multicolumn{1}{c||}{$0.099$} 
& \multicolumn{1}{c|}{$0.123$} & \multicolumn{1}{c|}{$0.099$} \\
\hline
$\chidof$ & \multicolumn{2}{c||}{$0$} & 
\multicolumn{1}{c|}{$9.3/6$} & 
\multicolumn{1}{c|}{$0.5/6$} \\
\hline \hline
$\De x^{\mathrm{exp}}/10^{-3}$     
& \multicolumn{2}{r||}{$10.1\pm4.2$\hspace*{3.5em}}
&  $9.9\pm4.2$ & $10.1\pm4.2$ \\
\hline
$\De y^{\mathrm{exp}}/10^{-3}$     
& \multicolumn{2}{r||}{$5.4\pm4.3$\hspace*{3.5em}}
& $7.0\pm4.3$ & $5.7\pm4.3$ \\
\hline
$\eps^{\mathrm{exp}}/10^{-3}$      
& \multicolumn{2}{r||}{$-5.3\pm1.6$\hspace*{3.5em}}
& $-4.2\pm1.5$ & $-5.0\pm1.5$ \\
\hline
$\De\yb^{\mathrm{exp}}/10^{-3}$    
& $-14.6\pm6.9$ & $-20.9\pm7.0$ 
&   $0.7\pm4.7$ & $-20.5\pm4.7$ \\
\hline
$\De\yh^{\mathrm{exp}}/10^{-3}$    
& $4.9\pm2.3$ &  $-1.7\pm2.3$ & 
\multicolumn{2}{c|}{$-1.4\;$ (from theory)} \\
\hline
$\De y_\nu^{\mathrm{exp}}/10^{-3}$ 
& \multicolumn{2}{r||}{$0.6\pm5.2$\hspace*{3.5em}}
& \multicolumn{2}{c|}{$-3.0\;$ (from theory)} \\
\hline
\end{tabular}
\ece
\caption[xxx]{Experimental results for the effective parameters 
for $\alpsz=0.123$ and the low value $\alpsz=0.099$.
The entries on the left-hand side are obtained by determining the six
effective parameters from the six observables
$\Gl$, $\swbar^2(\LEP)$, $\MW$, $\GT$, $\Gh$, and $\Gb$. On the
right-hand side $\De\yh$ and $\De y_\nu$ have been taken from theory, and
the remaining parameters have been fitted twice, namely for fixed
$\alpsz=0.123$ and with $\alpsz$ as additional fit parameter, resulting
in $\alpsz=0.099$.}
\label{tab:effpar}
\etab

As a final point, we compare the $(\Mt,\MH,\De\yb,\alpz,\alpsz)$-fit 
discussed above
with an analysis based on {\it phenomenological effective parameters}.
The six observables $\Gl$, $\swbar^2(\LEP)$, $\MW$ and $\GT$, $\Gh$, $\Gb$ can
be represented as linear combinations of six phenomenological parameters 
$\Delta x$, $\Delta y$, $\varepsilon$ and 
$\Delta\yh=(\Delta\yu+\Delta\yd)/2+(s^2_0/6c^2_0)(\Delta\yd-\Delta\yu)$,
$\Delta\yb$, $\Delta y_\nu$ that describe
possible sources of SU(2) violation within
an effective Lagrangian for electroweak interactions at the Z-boson 
resonance~\cite{zph2}.  We assume that the QCD corrections, 
such as $R_\QCD$, which enter $\GT$, $\Gh$, 
and $\Gb$, have standard form. These corrections are 
extracted from the experimental data before the determination of the 
effective parameters (see \citere{zph2}), which therefore quantify
all electroweak corrections 
to the $\alpz$-Born approximation.

The results of extracting the experimental values of
the six parameters $\Delta x^{\exp}$ etc.\ from the six
observables by inverting the system of linear 
equations is shown on the left-hand side of \refta{tab:effpar}.
The $\alps$-dependence
only affects $\Delta\yb^{\exp}$ and $\Delta\yh^{\exp}$, 
since the leptonic sector by itself determines $\Delta x^{\exp}$, 
$\Delta y^{\exp}$ and $\varepsilon^{\exp}$. 
The values of $\Delta x^{\exp}$, $\Delta y^{\exp}$ and
$\varepsilon^{\exp}$ are in excellent agreement with their SM
predictions, as discussed 
in detail for the data of \citeres{data8/95,SLD} in
\citere{zph3}. For $\alpsz=0.123$, in addition to the
non-standard value of $\De\yb^{\exp}$, also the parameter $\De\yh^{\exp}$
disagrees with the theoretical prediction \cite{zph2} of 
$\Delta\yh^{\SM}  = - 3.0 \times 10^{-3}$. Agreement between SM and
experiment in $\Delta\yh$ is achieved, however, for 
low values of $\alpsz$, such as $\alpsz=0.099$.

Noting that the process-specific parameters $\De\yh$ and $\De y_\nu$
only depend \cite{zph2} on the 
empirically well-established%
\footnote{Here we ignore the $\Rc$ problem, previously commented upon.}
couplings between vector-bosons and light fermions (i.e.\ all fermions 
except for top and bottom quarks), we now impose the SM values for
$\De\yh$ and $\De y_\nu$ and determine the remaining parameters in a
fit. According to the right-hand side of \refta{tab:effpar}, for fixed
$\alpsz=0.123$, we find a rather poor quality of the fit ($\chidof=9.3/6$).
Allowing for $\alpsz$ as additional fit parameter, we obtain an
excellent quality of the fit ($\chidof=0.5/6$), and for $\alpsz$ the low
value of $\alpsz=0.099$ deliberately chosen before.

It has thus been shown that the low value for $\alpsz$, as a
consequence of the experimental value of $\Rb$, emerges independently of
much of the details of electroweak radiative corrections. The 
two very weak assumptions made here, namely standard
form of the QCD corrections and of $\Delta y_\nu$ and $\Delta\yh$,
already imply the very low value of $\alpsz = 0.099\pm0.008$
in the fit which includes $\Rb$. 
This value is very close to the one obtained above
in the $(\Mt,\MH,\De\yb,\alpz,\alpsz)$ fit, 
where non-standard contributions have only been allowed in the \Zbb\ vertex. 
Moreover, the values of $\Delta\yb = 0.7 \pm 4.7$ 
for $\alpsz = 0.123$ and of $\Delta\yb = -20.5\pm4.7$ 
for $\alpsz = 0.099$ 
(as well as $\De x^{\exp}$, $\De y^{\exp}$, $\eps^{\exp}$)
obtained in the present analysis
are in good agreement with the values given in \refta{tab:dyb} for the
$(\Mt,\MH,\Delta\yb,\alpz)$ fit. 
In both treatments a decent value of $\chidof$
therefore requires a value of $\alpsz$
that, in the best-fit case, is four standard deviations below 
$\alpsz = 0.123\pm0.006$.

In summary, from our analysis of the precision data at the Z-boson
resonance and $\MW$, we find that a Higgs-boson mass lying in the
perturbative regime of the 
Standard Model, i.e.\ below $1\TeV$, is indeed favored
at the $1\si$ level.
Having investigated in much detail the impact of the data 
for the \Zbb,$\Pc\bar\Pc$ decay modes and 
the experimental value of $\swbar^2(\SLD)$, as well as the 
influence of the uncertainties connected with the input
parameters $\alpz$ and $\alpsz$, we conclude that a stronger upper
1$\sigma$ bound on $\MH$ than
$\MH\lsim900\GeV$ based on $\swbar^2(\LEP)$ and 
$\MH\lsim430\GeV$ based on $\swbar^2(\LEP + \SLD)$ can hardly be justified 
from the data at present. The stringent bounds on $\MH$ that are obtained when
the unmodified 
Standard Model is fitted to the complete data sample are immediately
lost when $\Rb$ and $\swbar^2(\SLD)$ are excluded from the analysis or,
as demonstrated for the case of $\Rb$, if non-standard contributions are
allowed in the theoretical model. The well-known fact that allowing
for a non-standard contribution to $\Rb$ 
gives rise to an
extremely low value of $\alpsz$ has been shown to emerge already under the 
weak theoretical assumptions of standard QCD corrections and standard form of
the couplings of the gauge-bosons to the 
leptons and to the quarks of the first two generations.

\vspace*{1em}
{\bf Note added in proof:} 
The most recent value from the Tevatron on $\Mt$ is
given by $\Mt^{\exp}=175\pm9\GeV$. All essential conclusions of
the present work, based on $\Mt^{\exp}=180\pm12\GeV$, remain valid
if this most recent value of $\Mt^{\exp}$ is used.

\section*{Acknowledgement}
G.W.\ thanks A.\ Djouadi for useful discussions.

\clearpage

\begin{figure}
\begin{center}
\begin{picture}(16,13.5)
\put(-2.2,-13.0){\includegraphics{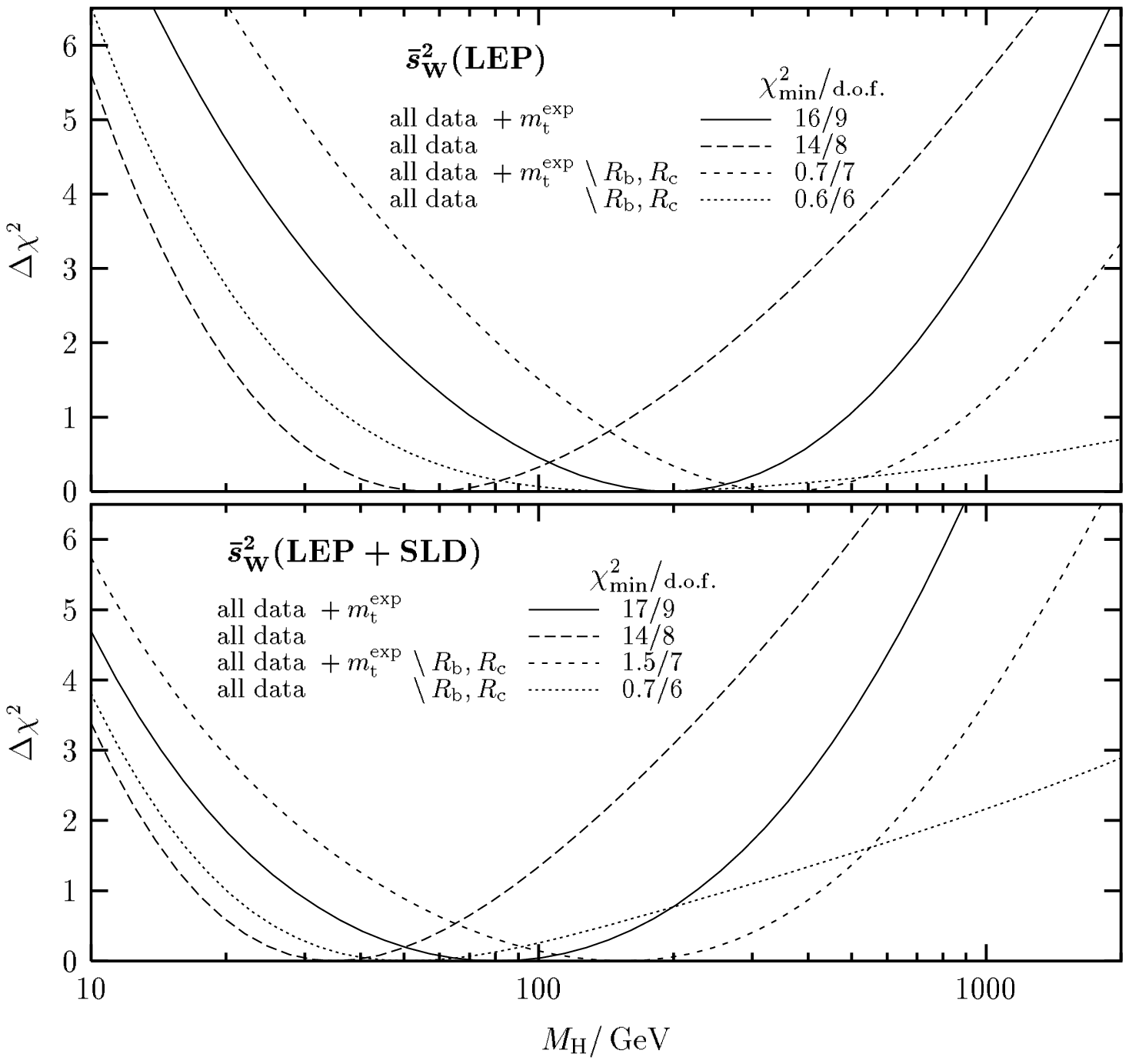}}
\end{picture}
\end{center}
\caption{$\De\chi^2=\chi^2-\chi^2_{\mathrm{min}}$ is plotted against
$\MH$ for the $(\Mt,\MH,\alpz,\alpsz)$ fit to various sets of 
physical observables, as specified in \refta{tab:globfit}.}
\label{fig:Dchi}
\efi

\clearpage

\begin{figure}[b]
\caption[xxx]{
The results of the two-parameter $(\Mt,\MH)$ fits within the SM
are displayed in the $(\Mt, \MH)$ plane. The three different columns refer 
to the three different sets of experimental data used in the corresponding 
fits, \protect\\
(i) ``leptonic sector'': $\Gl,\swbar^2(\LEP),\MW$, \protect\\
(ii) ``all data $\backslash\Rc,\Rb$'': $\GT,\Gh$ are added to set (i),
\protect\\
(iii) ``all data $\backslash\Rc$'': $\GT,\Gh,\Gb$ are added to the set (i).
\protect\\
The second and the third rows show the shift resulting from changing 
$\alpz^{-1}$ and $\alpsz$, respectively,
by one standard deviation in the SM prediction.
The fourth row shows the effect of replacing $\swbar^2(\LEP)$ by 
$\swbar^2(\SLD)$ and $\swbar^2(\LEP + \SLD)$ in the fits.
Note that the $1\sigma$ boundaries given in the first row 
are repeated identically in each row, in order to facilitate comparison with 
other boundaries. 
The value of $\chidof$ indicated in the plots refers to the 
central values of $\alpz^{-1}$ and $\alpsz$.
In all plots the empirical value of the top-quark mass
(not included as input of the fits)
of $\Mt^{\exp} = 180 \pm 12 \GeV$ is also indicated.
}
\label{fig:mtmhfit}
\efi

\clearpage

\addtocounter{figure}{-1}
\begin{figure}
\begin{center}
\begin{picture}(16,21)
\put(-3.2,-5.7){\includegraphics{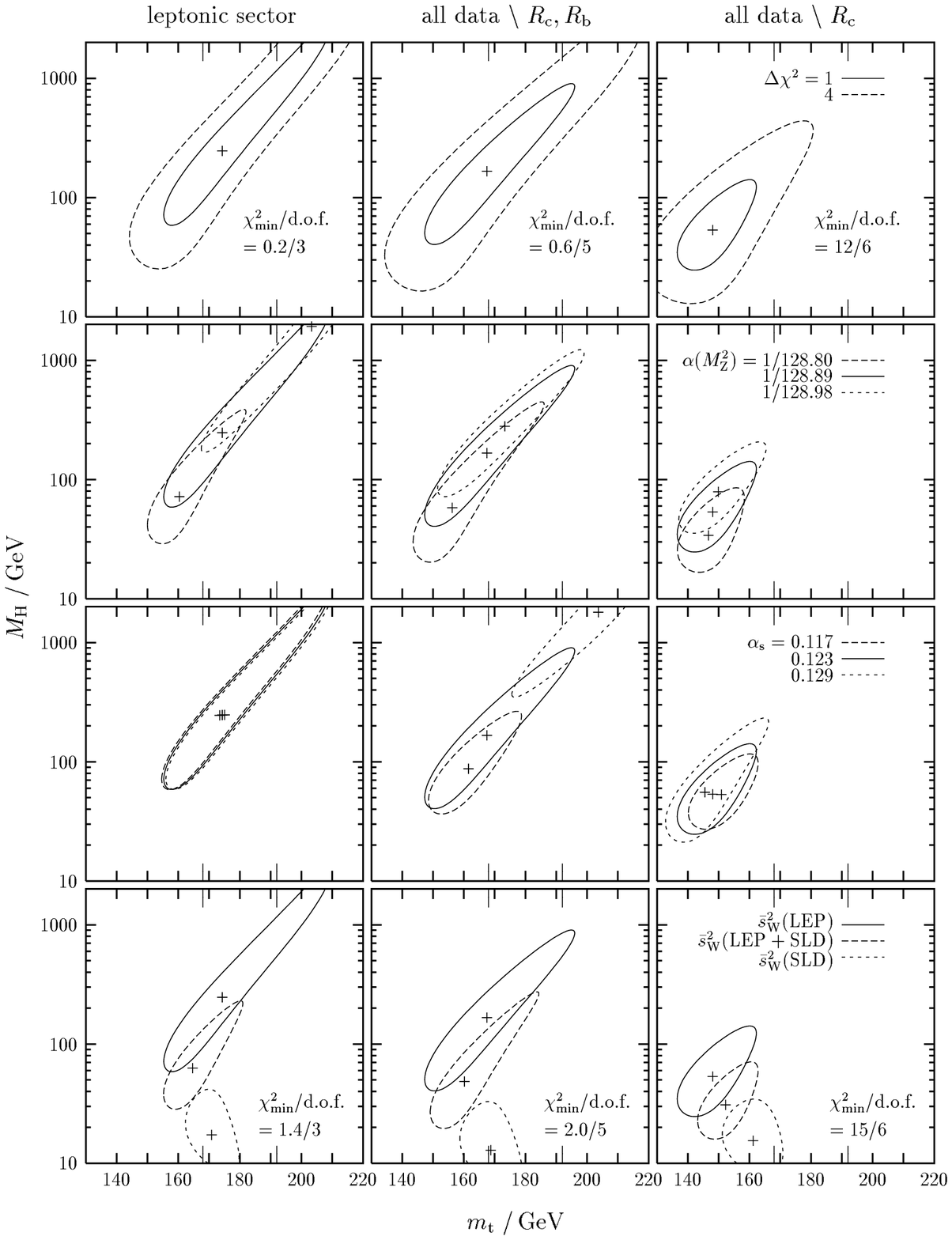}}
\end{picture}
\end{center}
\caption{}
\efi

\end{document}